\documentclass{article}
\usepackage{spconf,amsmath,graphicx,hyperref,booktabs,anyfontsize}
\usepackage{enumitem,float,mathtools,amssymb,caption,multirow,float}
\usepackage{siunitx}
\usepackage[activate={true,nocompatibility},final,tracking=true,kerning=true,spacing=true,factor=1100,stretch=10,shrink=10]{microtype}
\title{FCPE: A Fast Context-based Pitch Estimation Model}
\name{Yuxin Luo$^{1,\dagger}$, Ruoyi Zhang$^{1,\dagger}$\thanks{$^{\dagger}$These authors contributed equally to this work.}, Lu-Chuan Liu$^2$, Tianyu Li$^1$, Hangyu Liu$^{1,*}$\thanks{$^*$Corresponding Author.}}
\address{$^1$Fish Audio, Santa Clara, CA, USA\\ $^2$University of Science and Technology of China, Hefei, Anhui, China}
\begin{document}
\ninept
\maketitle
\begin{abstract}
Pitch estimation (PE) in monophonic audio is crucial for MIDI transcription and singing voice conversion (SVC), but existing methods suffer significant performance degradation under noise. In this paper, we propose FCPE, a fast context-based pitch estimation model that employs a Lynx-Net architecture with depth-wise separable convolutions to effectively capture mel spectrogram features while maintaining low computational cost and robust noise tolerance. Experiments show that our method achieves 96.79\% Raw Pitch Accuracy (RPA) on the MIR-1K dataset, on par with the state-of-the-art methods. The Real-Time Factor (RTF) is 0.0062 on a single RTX 4090 GPU, which significantly outperforms existing algorithms in efficiency. Code is available at \href{https://github.com/CNChTu/FCPE}{https://github.com/CNChTu/FCPE}.
\end{abstract}

\section{Introduction}
\label{sec:intro}

Pitch estimation (PE) or fundamental frequency (f0) estimation is crucial for tasks such as MIDI transcription and singing voice conversion (SVC). Vocal pitch estimation is one of the cases that is widely used for industrial production. Before the advent of recent deep learning models, the field of PE was mainly dominated by classical signal processing techniques. These methods can be categorized into three types.

\begin{itemize}[wide, labelwidth=!, labelindent=0pt, itemsep=0pt, parsep=0pt]
    \item \textbf{Time-domain methods:} They operate directly on the signal waveform to identify its periodic structure. The most well-known one is the Autocorrelation Function (ACF) \cite{1162846} method.
    \item \textbf{Frequency-domain methods:} These methods transform the signal into the frequency domain to analyze its harmonic structure. The cepstrum-based pitch determination algorithm \cite{Noll1967} is a classic example designed to separate the glottal source excitation from the vocal tract filter to reveal the fundamental frequency.
    \item \textbf{Hybrid methods:} To achieve greater robustness, these methods combine techniques from both domains. The seminal YIN algorithm \cite{de2002yin}, for instance, enhances the ACF by incorporating several error-reduction steps, while YAAPT \cite{zahorian2011yaapt} explicitly integrates spectral harmonic information with time-domain analysis to improve tracking accuracy.
\end{itemize}

The methods above represent the traditional paradigm of PE, while achieving considerable success, they still struggle with noisy environments, polyphonic sources, and other issues.

With the development of deep learning (both software and hardware), related work has greatly enhanced results in PE tasks. Seminal models such as CREPE \cite{kim2018crepe}, DeepF0 \cite{singh2021deepf0} and HARMOF0 \cite{wei2022harmof0} have leveraged Convolutional Neural Network (CNN) or Recurrent Neural Network (RNN) to establish new state-of-the-art benchmarks for accuracy and robustness. More recently, the introduction of RMVPE \cite{wei2023rmvpe} marked another significant milestone by adapting newer models like U-Net to reach an unprecedented performance. However, current deep learning methods require substantial computation and introduce latency, limiting their real-time applications. For RMVPE, in particular, its high computational demand is a direct consequence of its sophisticated architecture.

To address these issues, we propose \textbf{FCPE (Fast Context-based Pitch Estimation)}, a novel model achieving high efficiency without sacrificing accuracy. FCPE employs a Lynx-Net backbone \cite{yxlllc2023lynxnet}, which leverages depthwise separable convolutions to efficiently extract features from mel spectrograms and offers sufficient contextual coverage to model temporal relationships across frames. In this paper, we demonstrate that this architectural choice, combined with elaborated training strategies, jointly enables our model to achieve state-of-the-art performance.

\section{FCPE}
\label{sec:pagestyle}

\subsection{Overall architecture}

We define the model input as \( X_{T \times F} \), where \(X\) represents the log mel-spectrogram, \(T\) is the number of mel frames, and \(F\) is the number of logarithmically-spaced frequency bins. Our task can be formulated as \(F\): \( X_{T \times 128} \) → \(Y_{T \times 360} \), \(Y\) is the predicted pitch matrix. The overall structure is shown in Figure \ref{fig:enter-label} and can be broken down into three stages:

\begin{figure}[h]
    \centering
    \includegraphics[width=0.75\linewidth,height=4.4cm]{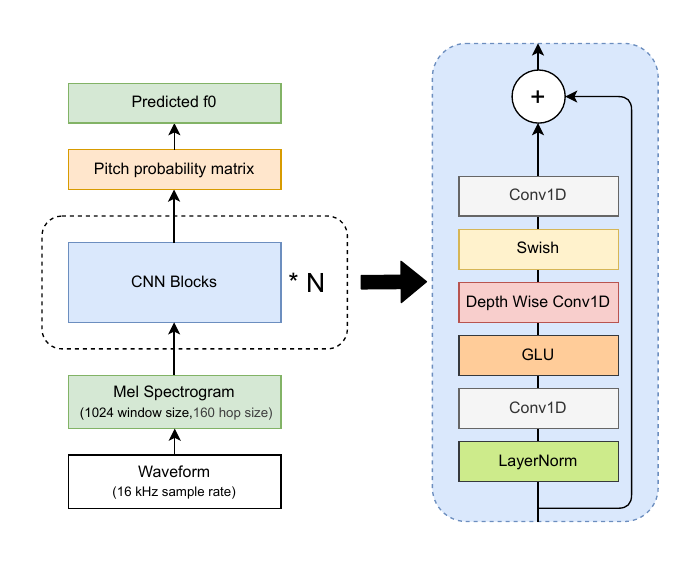}
    \caption{Overall architecture of FCPE.}
    \label{fig:enter-label}
    \vspace{-10pt}  % 减少图片后的垂直间距
\end{figure}

\begin{itemize}[wide, labelwidth=!, labelindent=0pt, itemsep=3pt, parsep=0pt]
    \item \textbf{Input Representation}: As depicted in Figure 1, the process begins with the raw audio waveform, which is first transformed into a log-mel spectrogram. This spectral representation is then passed through an initial embedding block. This block, composed of shallow 1D convolutional layers, maps the input features into a high-dimensional vector sequence suitable for the main backbone. Optionally, a learnable harmonic embedding can be added to this sequence to explicitly enhance harmonic features before it is fed into the next stage.
    \item \textbf{Lynx-Net Backbone}: The core of FCPE is a stack of lightweight Lynx-Net layers (CNN Blocks in Fig. 1), designed for efficient temporal context modeling. As shown in Figure \ref{fig:enter-label}, each block uses a Conformer-inspired \cite{gulati2020conformerconvolutionaugmentedtransformerspeech} structure. Its key component is a depthwise Conv1D layer, which efficiently captures local patterns. Pointwise convolutions manage channel dimensions, and a residual connection facilitates the training of deeper networks.
    \item \textbf{Output Stage}: After passing through the Lynx-Net layers, the refined feature sequence is projected by a final linear layer to produce the pitch probability matrix \(Y_{T \times 360} \). This matrix represents the probability of each cent bin for every time frame. To derive the final f0, we do not simply take the bin with the highest probability. Instead, we apply a local argmax function \cite{kim2018crepe,wei2023rmvpe} as a decoding step. This function calculates a weighted average of pitches around the peak probability bin, providing a more precise and robust f0 estimate than a simple argmax operation, thereby generating the final predicted f0.
\end{itemize}

\subsection{Loss function and decoding strategy}

Following previous works \cite{kim2018crepe,wei2023rmvpe}, we formulate pitch estimation as a classification problem over 360 discrete pitch bins. Each bin corresponds to a specific pitch value defined in cents relative to $f_{\text{ref}} = 10$ Hz:

\vspace{-8pt}
\begin{equation}
    c(f) = 1200 \cdot \log_2 \frac{f}{f_{\text{ref}}}
\end{equation}

The 360 pitch values are denoted as $c_1, c_2, \ldots, c_{360}$ and are selected to cover six octaves with 20-cent intervals between C1 and B7, corresponding to 32.70 Hz and 1975.5 Hz respectively. This logarithmic pitch scale yields 100 cents per semitone, enabling fine-grained pitch resolution.

The model outputs a probability vector $\hat{y}$ over these bins. We use the BCE loss for training:

\vspace{-12pt}
\begin{equation}
    \mathcal{L}(y, \hat{y}) = -\sum_{i=1}^{360} \left( y_i \log \hat{y}_i + (1 - y_i) \log(1 - \hat{y}_i) \right),
\end{equation}where the target $y$ is defined exactly as in the CREPE paper.

At inference, we employ a local weighted average decoding mechanism \cite{wei2023rmvpe} to obtain the final pitch estimate. If the confidence score exceeds a threshold, the pitch estimate $\hat{c}$ in cents is calculated as a weighted average around the peak probability:

\vspace{-8pt}
\begin{equation}
    \hat{c} = \sum_{i=m-4}^{m+4} (\hat{y}_i c_i) \bigg/ \sum_{i=m-4}^{m+4} \hat{y}_i, \quad m = \arg \max_i \hat{y}_i
\end{equation}

\vspace{-4pt}
\begin{equation}
    \hat{f} = \begin{cases} 
    f_{\text{ref}} \times 2^{\hat{c}/1200} & \text{if } \max_i \hat{y}_i \geq 0.05 \\
    0 & \text{otherwise}
    \end{cases}
\end{equation}

\vspace{-4pt}
\subsection{Training details}

\begin{figure}[h]
    \centering
    \includegraphics[width=0.75\linewidth]{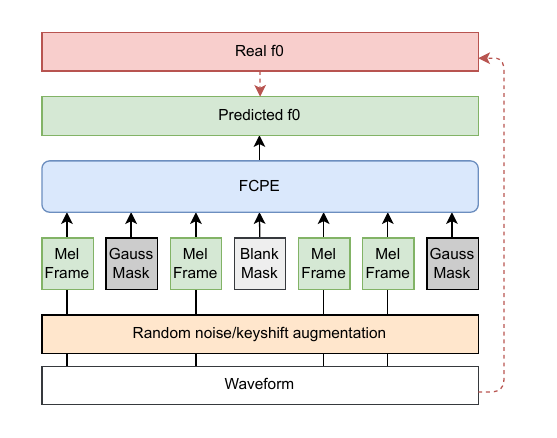}
    \caption{Details of training strategies.}
    \label{fig:placeholder}
\end{figure}

To establish an objective and precise ground truth and eliminate subjective errors from manual labeling, we re-synthesize the M4Singer \cite{zhang2022msinger} and VCTK \cite{https://doi.org/10.7488/ds/2645} datasets using the Differentiable Digital Signal Processing (DDSP) \cite{engel2020ddsp} method and use them for training all model variants. We then employ several data augmentation strategies. First, we apply random key shifting to the raw audio waveforms to increase pitch diversity. Next, to enhance the model's noise robustness, we superimpose various types of noise (white, colored, real-world). Finally, we apply random masking to the augmented mel-spectrograms, which, as depicted in Figure \ref{fig:placeholder}, involve either blank or Gaussian masking. This compels the model to infer pitch from the surrounding contextual information rather than relying on isolated features of a single frame. The effectiveness of these techniques is empirically validated in our experiments, showing significant improvements in model robustness, particularly in noisy conditions. 

\section{Experiments}

To compare FCPE with other models and quantify training strategies' contribution to the system, we conduct a series of experiments. The following parts illustrate the results of the comparative study and ablation study. 

\begin{table*}[h!]
\centering

\caption{Performance comparsion of six algorithms. One that do not list the number of parameters are traditional signal processing algorithms.}

\label{tab:sota_comparison_combined}
\setlength{\tabcolsep}{2pt}
\renewcommand{\arraystretch}{1.1}
\begin{tabular}{lcccc|cccc|cccc|cccc}
\toprule
& \multicolumn{4}{c|}{\textbf{MIR-1K dataset}} & \multicolumn{4}{c|}{\textbf{THCHS30-Synth dataset}} & \multicolumn{4}{c|}{\textbf{Vocadito dataset}} & \multicolumn{4}{c}{\textbf{TONAS dataset}} \\
\cmidrule(lr){2-5} \cmidrule(lr){6-9} \cmidrule(lr){10-13} \cmidrule(lr){14-17}
\textbf{Algorithm} & \multicolumn{4}{c|}{\textbf{RPA (\%) ↑}} & \multicolumn{4}{c|}{\textbf{RPA (\%) ↑}} & \multicolumn{4}{c|}{\textbf{RPA (\%) ↑}} & \multicolumn{4}{c}{\textbf{RPA (\%) ↑}} \\
\cmidrule(lr){2-5} \cmidrule(lr){6-9} \cmidrule(lr){10-13} \cmidrule(lr){14-17}
(Parameters) & \textbf{Clean} & \textbf{20 dB} & \textbf{0 dB} & \textbf{-20 dB} & \textbf{Clean} & \textbf{20 dB} & \textbf{0 dB} & \textbf{-20 dB} & \textbf{Clean} & \textbf{20 dB} & \textbf{0 dB} & \textbf{-20 dB} & \textbf{Clean} & \textbf{20 dB} & \textbf{0 dB} & \textbf{-20 dB} \\
\midrule
\multicolumn{17}{l}{\textbf{White Noise ($\beta$=0)}} \\
\midrule
FCPE (10.64M) & 96.79 & 97.06 & 97.09 & 29.75 & \textbf{97.56}& \textbf{96.48}& 81.44& \textbf{12.17}& 95.80 & 95.82 & 93.43 & 21.60 & \textbf{96.03} & \textbf{95.86} & 64.56 & \textbf{4.69} 
\\
RMVPE (90.42M) & 97.77 & 97.57 & \textbf{97.39} & \textbf{43.63} & 96.37& 95.79& \textbf{84.12}& 3.98& 95.83 & 95.81 & \textbf{95.47} & \textbf{28.80} & 95.64 & 95.66 & \textbf{81.49} & 1.82 
\\
CREPE (22.24M) & 97.90 & 97.89 & 94.07 & 1.09 & 89.67& 88.81& 65.78& 0.16& \textbf{97.35} & \textbf{97.35} & 93.08 & 0.56 & 95.20 & 95.20 & 79.69 & 0.02 
\\
PESTO (0.13M) & \textbf{98.47} & \textbf{98.38} & 95.48 & 17.78 & 89.72& 86.47& 60.59& 7.01& 95.50 & 95.24 & 89.18 & 13.20 & 91.30 & 89.31 & 64.59 & 4.10 
\\
PM & 96.06 & 95.63 & 20.90 & 0.00 & 77.57& 77.52& 14.14& 0.00& 91.75 & 91.62 & 19.27 & 0.00 & 89.96 & 88.83 & 10.81 & 0.00 
\\
Harvest & 95.11 & 95.03 & 62.02 & 0.17 & 87.93& 87.06& 40.70& 0.05& 94.17 & 93.88 & 72.19 & 0.17 & 93.98 & 93.00 & 11.27 & 0.01 \\
\midrule
\multicolumn{17}{l}{\textbf{Pink Noise ($\beta$=1)}} \\
\midrule
FCPE (10.64M) & 96.79 & 97.12 & 95.97 & \textbf{18.04} & \textbf{97.56}& \textbf{96.30}& 76.42& \textbf{8.40}& 95.80 & \textbf{95.88} & 92.28 & \textbf{15.65} & \textbf{96.03} & \textbf{95.81} & 61.74 & \textbf{4.85} 
\\
RMVPE (90.42M) & 97.77 & 97.59 & \textbf{96.61} & 13.66 & 96.37& 95.34& \textbf{77.46}& 2.83& 95.83 & 95.84 & \textbf{94.82} & 9.15 & 95.64 & 95.72 & \textbf{77.29} & 2.41 
\\
CREPE (22.24M) & 97.90 & 97.79 & 91.72 & 2.54 & 89.67& 87.61& 59.03& 0.62& \textbf{97.35} & 97.27 & 90.92 & 1.91 & 95.20 & 94.66 & 73.05 & 0.25 
\\
PESTO (0.13M) & \textbf{98.47} & \textbf{98.39} & 92.62 & 10.24 & 89.72& 86.34& 61.97& 6.68& 95.50 & 95.16 & 88.09 & 9.90 & 91.30 & 90.05 & 67.42 & 3.47 
\\
PM & 96.06 & 96.02 & 49.84 & 0.00 & 77.57& 77.89& 26.66& 0.00& 91.75 & 91.78 & 52.99 & 0.00 & 89.96 & 89.61 & 42.76 & 0.00 
\\
Harvest & 95.11 & 94.98 & 51.06 & 0.06 & 87.93& 85.89& 27.81& 0.04& 94.17 & 93.68 & 62.04 & 0.10 & 93.98 & 92.52 & 9.25 & 0.01 \\
\midrule
\multicolumn{17}{l}{\textbf{Real-World Noise (CHiME)}} \\
\midrule
FCPE (10.64M) & 96.78 & 96.95 & 90.59 & 35.83 & \textbf{97.56}& \textbf{96.47}& 80.51& 23.87& 95.80 & \textbf{95.87}& 91.17 & 38.03 & \textbf{96.03} & \textbf{95.88} & 76.83 & 15.64 \\
RMVPE (90.42M) & 97.77 & 97.89 & \textbf{95.13} & \textbf{41.53} & 96.37
& 95.56& \textbf{84.37}& \textbf{30.81}& 95.83 & 95.85& \textbf{93.71} & \textbf{43.72} & 95.64 & 95.62 & \textbf{86.56} & \textbf{30.86} \\
CREPE (22.24M) & 97.90 & 97.70 & 91.43 & 30.20 & 89.67
& 87.64& 66.37& 13.66& \textbf{97.35} & 97.24 & 91.16 & 31.62 & 95.20 & 94.54 & 76.44 & 13.84 \\
PESTO (0.13M) & \textbf{98.47} & \textbf{98.34} & 91.14 & 35.84 & 89.72
& 88.41& 71.42& 23.74& 95.50 & 95.38 & 89.24 & 38.43 & 91.30 & 90.33 & 73.32 & 23.54 \\
PM & 96.06 & 95.39 & 62.91 & 4.15 & 77.57
& 76.70& 44.86& 3.41& 91.75 & 91.36 & 65.89 & 2.96 & 89.96 & 89.27 & 63.02 & 3.43 \\
Harvest & 95.11 & 94.87 & 77.59 & 12.85 & 87.93& 86.24& 58.17& 8.35& 94.17 & 93.81 & 80.36 & 15.89 & 93.98 & 92.15 & 56.58 & 6.55 \\
\bottomrule
\end{tabular}%
\end{table*}

\vspace{-8pt}
\subsection{Comparative study} % PVP环节

\subsubsection{Accuracy} 

In order to demonstrate the effectiveness and robustness of our model, we conduct a comparative study against five leading pitch estimation algorithms: RMVPE, CREPE, PESTO \cite{PESTO}, PM \cite{2000Praat}, and Harvest \cite{morise17b_interspeech}. The performance is evaluated on the MIR-1K \cite{hsu2009improvement,raffel2014mir_eval}, Vocadito \cite{bittner_2021_5578807}, and TONAS \cite{cofla_computational_analysis_of_flamenco_2013_1290722} datasets under various challenging conditions, including clean audio and audio corrupted by colored noise and real-world environmental noise from the CHiME dataset \cite{foster2015chime} at multiple Signal-to-Noise Ratios (SNRs). Additionally, to reduce the manual labeling's errors, we also use another DDSP re-synthesized dataset in the evaluation. Specifically, it is re-synthesized from the test set of the THCHS30 \cite{THCHS30_2015} dataset, and we refer to this dataset as THCHS30-Synth. In this comparison, all deep learning-based models are evaluated using their publicly available pre-trained versions \cite{pitchtracker}. It should be noted that since the MIR-1K dataset was used in the training of RMVPE, CREPE, and PESTO, their superior performance on this specific dataset is expected. The four datasets contain various vocal data that cover a wide range from singing to speech.

The Raw Pitch Accuracy (RPA) is presented in Table \ref{tab:sota_comparison_combined}. All evaluations are conducted five times and the average values are reported in the table. As shown in the table, our model achieves remarkable performance across multiple datasets, even in the presence of severe noise interference, it still demonstrated excellent robustness. The results clearly indicate that FCPE's accuracy is highly competitive with the state-of-the-art robust model, RMVPE. This consistent, high-level performance validates our approach, proving that FCPE delivers exceptional robustness and accuracy for a wide range of scenarios, making it a powerful and efficient tool for pitch estimation.

Another aspect is that our model doesn't have as many parameters as RMVPE or CREPE, but we still achieve performance that is comparable to or even surpasses theirs, which fully demonstrates the effectiveness of our model architecture and training strategy.

\subsubsection{Real-time factor} 

In addition to accuracy, computational efficiency is a critical factor for real-world applications such as real-time singing voice conversion. To evaluate the inference speed of FCPE, we measure its Real-Time Factor (RTF). The RTF is a standard metric for assessing the efficiency of audio processing models, defined as the time required to process an audio stream divided by the duration of the audio itself. A model with an RTF significantly lower than 1.0 is considered to be capable of real-time processing. The RTF is calculated as follows:

\begin{equation}
\label{eq:rtf}
\text{RTF} = \frac{T_{\text{process}}}{T_{\text{audio}}}
\end{equation}
where $T_{\text{process}}$ is the time taken by the model to infer the pitch for an audio clip of duration $T_{\text{audio}}$.

We also calculate the FLOPS (Floating Point Operations) needed to process one-second audio, lower FLOPS requirements also indicate lower inference costs.

\begin{table}[!htbp]
\centering
\caption{Real-Time Factor comparison and FLOPS needed for inference one second of different pitch estimation models. The test is run on a single RTX 4090 GPU.}
\label{tab:rtf_comparison}
\begin{tabular}{lcc}
\toprule
\textbf{Model} & \textbf{RTF} & \textbf{FLOPS} \\
\midrule
\textbf{FCPE (ours)} & \textbf{0.0062} & \textbf{1.06 GFLOPS}\\
PESTO (MIR-1K\_g7)& 0.0164 & 2.82 GFLOPS\\
RMVPE& 0.0329 & 4.91 GFLOPS\\
CREPE          & 0.4775 & 141 GFLOPS\\
\bottomrule
\end{tabular}
\end{table}

The results, presented in Table~\ref{tab:rtf_comparison}, unequivocally demonstrate the computational efficiency of FCPE. Our model achieves an RTF of 0.0062, which is approximately \textbf{5.3x} faster than RMVPE, \textbf{2.6x} faster than PESTO even though its parameters is far less than ours, and an astounding \textbf{77x} faster than the widely-used CREPE model.

\vspace{3pt}

The significant speed advantage is a direct result of FCPE's lightweight architecture, which heavily relies on efficient depthwise separable convolutions. This not only confirms FCPE's suitability for real-time applications, such as live voice conversion and music information retrieval, but also makes it an excellent candidate for large-scale batch processing and deployment on resource-constrained edge devices.

\subsection{Ablation study}

% 消融实验都是在训练阶段做的效果，DDSP合成的数据作为训练集，然后在训练的过程中对数据进行随机增强（加噪，加mask，变调）
% 我们的消融是[全加](baseline),[去掉加噪],[去掉mask],[去掉变调];评估指标为[RCA][RPA][OA][VFA][VR],参数变量为[beta](有色噪声beta),[dB](加噪强度){99dB->信噪比无穷大}
% (先说结论，又快又robust) 消融实验的结果展示出如下结论：1.我们的训练方案显著地提高了模型的鲁棒性，使得它能够应对更加极端的条件。2.mask相关的训练方案展示出FCPE长上下文的特性，能够很好地捕捉前后关联，抵抗极端噪声。3.因为可用数据过少，为了覆盖更广的音域，我们不得不采用变调增强的方案，根据我们的实验发现，该方案会导致模型的准确率略微地下降。
% 综上所述，我们的训练方案，能够有效地利用模型长上下文的特性，从而提升算法的性能及鲁棒性。

To investigate the individual contribution of each training strategy to the model's performance, we conduct a series of ablation studies. Our model is trained with a combination of three key data augmentation techniques: \textbf{noise augmentation}, \textbf{spectrogram masking}, and \textbf{random key shifting}. We then create variants of the model by systematically removing each of these components during training.

\begin{table*}[h]
\centering
\caption{ Ablation results of different training strategies organized by noise type on the MIR-1K dataset. (↑: Higher is better).}
\label{tab:ablation_reorganized_rpa_expanded}
\setlength{\tabcolsep}{2pt}
\renewcommand{\arraystretch}{1.6}
\begin{tabular}{l|cccc|cccc|cccc|cccc}
\toprule
\multirow{2}{*}{\textbf{Strategy}} & \multicolumn{4}{c|}{\textbf{White Noise ($\boldsymbol{\beta}=0$)}} & \multicolumn{4}{c|}{\textbf{Pink Noise ($\boldsymbol{\beta}=1$)}} & \multicolumn{4}{c|}{\textbf{Brownian Noise ($\boldsymbol{\beta}=2$)}} & \multicolumn{4}{c}{\textbf{Violet Noise ($\boldsymbol{\beta}=-1$)}} \\
\cmidrule(lr){2-5} \cmidrule(lr){6-9} \cmidrule(lr){10-13} \cmidrule(lr){14-17}
& \textbf{Clean} & \textbf{20 dB} & \textbf{0 dB} & \textbf{-20 dB} & \textbf{Clean} & \textbf{20 dB} & \textbf{0 dB} & \textbf{-20 dB} & \textbf{Clean} & \textbf{20 dB} & \textbf{0 dB} & \textbf{-20 dB} & \textbf{Clean} & \textbf{20 dB} & \textbf{0 dB} & \textbf{-20 dB} \\
\midrule
\textbf{Full Strategy} & 96.79 & 97.06 & 97.09 & 29.75 & 96.79 & 97.12 & 95.97 & \textbf{18.04} & 96.79 & 96.91 & 96.99 & 96.31 & 96.79 & 96.88 & 97.09 & 41.80 \\
\midrule
\textbf{w/o Noise Aug.} & 96.69 & 96.88 & 94.35 & 6.19 & 96.69 & 96.99 & 91.75 & 6.00 & 96.69 & 96.87 & 97.03 & 92.46 & 96.69 & 96.79 & 95.71 & 18.11 \\
\midrule
\textbf{w/o Spec. Mask.} & 96.58 & 96.87 & 96.97 & 16.72 & 96.58 & 96.93 & 95.64 & 4.81 & 96.58 & 96.66 & 96.75 & 96.00 & 96.58 & 96.72 & 97.03 & 55.41 \\
\midrule
\textbf{w/o Key Shift.} & \textbf{96.94} & \textbf{97.18} & \textbf{97.20} & \textbf{32.77} & \textbf{96.94} & \textbf{97.23} & \textbf{96.02} & 9.14 & \textbf{96.94} & \textbf{97.05} & \textbf{97.14} & \textbf{96.61} & \textbf{96.94} & \textbf{97.08} & \textbf{97.37} & \textbf{66.47} \\
\bottomrule
\end{tabular}%
\end{table*}

All model variants are trained on the same re-synthesized datasets mentioned in 2.3 and evaluated on the MIR-1K dataset. To rigorously assess robustness, the evaluation is conducted under a spectrum of synthetic noisy conditions, where we vary both the noise color (controlled by the $\beta$ parameter) and the signal-to-noise ratio (SNR in dB). We use two metrics for evaluation: RPA and RCA. The detailed results are presented in Table \ref{tab:sota_comparison_combined} below. The experiments are conducted five times and the average values are reported in the table (same as comparative study).

\vspace{2pt}

The results in Table \ref{tab:ablation_reorganized_rpa_expanded} offer a clear picture of each component's contribution. First, it is evident that the model trained without \textbf{noise augmentation} suffers a dramatic loss in accuracy under noisy conditions. This is most apparent in the extreme -20 dB white noise ($\beta$=0) scenario, where the RPA plummets from the baseline's \textbf{29.75\%} to a mere \textbf{6.19\%}. This stark degradation unequivocally demonstrates that noise augmentation is the most critical component for enhancing the model's robustness.

\vspace{2pt}

Second, the significance of \textbf{spectrogram masking} is also clearly demonstrated, especially against structured noise. With its removal, the model's RPA on -20 dB pink noise ($\beta$=1) drops to a mere \textbf{4.81\%}. Notably, this is even worse than the model without noise augmentation (\textbf{6.00\%}), which strongly supports our core hypothesis: masking is the key mechanism that compels FCPE to exploit its long-context architecture. This learned skill of finding reliable pitch cues from the temporal context is essential for overcoming extreme noise.

\vspace{2pt}

\begin{figure}
    \centering
    \includegraphics[width=1\linewidth, height=0.5\linewidth]{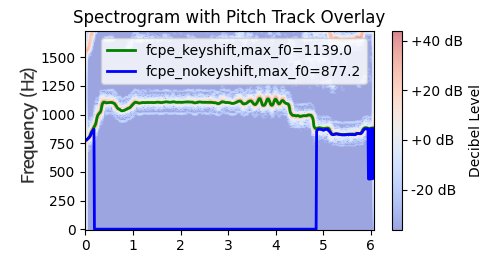}
    \setlength{\abovecaptionskip}{5pt}  % 减少caption上方间距
    \setlength{\belowcaptionskip}{-5pt} % 减少caption下方间距
    \caption{Model's vocal range with and without key shifting. The blue line incorrectly predicts no f0 in the middle section, while the green line (with key shifting) maintains continuous pitch tracking. }
    \label{fig:final_figure}
\end{figure}

Finally, removing \textbf{key shifting} reveals an interesting and important trade-off between model precision and generalization. At first glance, the model without this augmentation appears superior, achieving the highest accuracy on clean data across all variants. However, our test set (MIR-1K) cannot cover more general and realistic situations, and the inherent lack of pitch diversity in our synthesized training data necessitates key shifting to enhance generalization for real-world scenarios. As shown in Figure \ref{fig:final_figure}, the model trained without key shifting augmentation mistakenly identifies a high-pitched human voice segment as unvoiced (no f0), while the model trained with key shifting augmentation correctly detects the f0 in the same segment. Further quantitative analysis reveals that the model's vocal range with key shifting enhancement is \textbf{29.8\%} wider (1139 Hz vs 877.2 Hz).

\vspace{3pt}
In summary, the findings from the ablation study can be summarized as follows:

\begin{itemize}[wide, labelwidth=!, labelindent=0pt, itemsep=3pt, parsep=0pt]
    \vspace{3pt}
    \item \textbf{Noisy data is significant for model’s performance}: the FCPE model, which incorporates all augmentation techniques, achieves a strong and balanced performance across all metrics. This demonstrated that our proposed training strategy is effective at producing a robust and accurate pitch estimation model.
    \vspace{3pt}
    \item \textbf{Spectrogram mask enhances robustness}: experiments with and without spectrogram masking demonstrates that our model effectively captures temporal relationships between preceding and succeeding frames due to its long-context architecture, which exhibits robustness against extreme noise conditions.
    \vspace{3pt}
    \item \textbf{Key shifting expands vocal range:} at first glance, removing key shifting (w/o Key Shifting) led to a slight improvement in certain accuracy metrics. However, given the limited pitch range of our training data, this augmentation is a necessary trade-off. It enables the model to cover a wider musical range and improves generalization, which is critical for real-world applications.
\end{itemize}
% 感觉后面PVP实验出来可以直接用缩写了

\section{Conclusion}

In this paper, we propose FCPE, a fast context-based pitch estimation model that reaches state-of-the-art performance. We also introduce several training strategies to enhance model performance, which are proven effective through ablation studies. Additionally, we successfully address the data scarcity problem by utilizing DDSP-resynthesized data, and we find that even re-synthesized data demonstrates excellent generalization capabilities.Our future work will focus on applying FCPE to applications such as real-time SVC and other related tasks.

\bibliographystyle{IEEEbib}
\bibliography{refs}

\end{document}